\documentclass[twocolumn]{article}
\usepackage{amsmath}
\usepackage{abstract}
\usepackage{graphicx}
\usepackage{hyperref}

\usepackage{arxiv-twocolumn} 

\usepackage[utf8]{inputenc}
\usepackage{float}
\usepackage{placeins}

\usepackage{tikz}
\usepackage{enumitem}
\usepackage[english]{babel}

\setitemize{parsep=0pt, wide=3pt, leftmargin=*}
\setenumerate{parsep=0pt, wide=3pt, leftmargin=*}

\title{Self-Supervised Inference of Agents in Trustless Environments}

\author{
 Vladyslav Larin \\
  Fortytwo \\
  \texttt{vlarin@fortytwo.network} \\
  \And
  Ivan Nikitin \\
  Fortytwo \\
  \texttt{inikitin@fortytwo.network}
  \And
  Alexander Firsov \\
  Fortytwo \\
  \texttt{afirsov@fortytwo.network} \\
}

\begin{document}

\maketitle

\begin{abstract}
In this paper, we propose a novel approach where agents can form swarms to produce high-quality responses effectively. This is accomplished by utilizing agents capable of data inference and ranking, which can be effectively implemented using LLMs as response classifiers. We assess existing approaches for trustless agent inference, define our methodology, estimate practical parameters, and model various types of malicious agent attacks. Our method leverages the collective intelligence of swarms, ensuring robust and efficient decentralized AI inference with better accuracy, security, and reliability. We show that our approach is an order of magnitude faster than other trustless inference strategies reaching less than 125 ms validation latency.
\end{abstract}

\section{Introduction}
Decentralized AI inference represents a transformative shift in the deployment and utilization of artificial intelligence. Traditional AI systems rely heavily on centralized infrastructures, which can lead to computational bottlenecks and single points of failure. These centralized systems often face challenges related to scalability and adaptability in meeting the growing demand for AI inference.

To address these limitations, a distributed approach has emerged, spreading AI computation across multiple nodes in a network. This method enhances system resilience and enables more efficient resource utilization. In such decentralized systems, tasks such as model training, inference, and data processing are performed across a distributed network. Consequently, this eliminates the need for a central authority, allowing for greater flexibility and potentially reducing operational costs. \cite{sakshi2024}.

The evolution of decentralized AI has been facilitated by advances in blockchain technology, which provides a robust framework for trustless operations. Smart contracts enable the automation of service-level agreements (SLAs) and ensure that all parties adhere to predefined rules without the need for intermediaries. This is particularly beneficial in scenarios where trust and transparency are paramount, such as financial services and healthcare \cite{pocket2024}.

Several innovative frameworks and protocols have been developed to support decentralized AI inference. The AI Protocol's Decentralized Inference Clusters (DePIN) exemplify the shift from traditional centralized models to distributed frameworks, enabling efficient and scalable AI services by leveraging decentralized resources \cite{aiprotocol2024}. Similarly, platforms like Nesa offer decentralized query marketplaces that support high-security and privacy measures, facilitating the deployment of AI models in trustless environments \cite{nesa2024}.

Despite these advancements, significant challenges remain in achieving fast, trustless inference on large artificial neural networks (ANNs). The computational demands and latency associated with decentralized operations can hinder performance, especially for real-time applications. Additionally, ensuring the integrity and security of the inference process in the presence of malicious actors is a complex problem that requires robust solutions.

This paper explores the current state-of-the-art in decentralized AI inference, highlighting the advantages and limitations of various approaches. In light of existing challenges, we propose a novel method for self-supervised inference of agents in trustless environments, leveraging the collective intelligence of agent swarms to ensure high-quality responses. We demonstrate that Large Language Models (LLMs) can efficiently and quickly perform the response ranking necessary for swarm consensus. Additionally, our approach incorporates mechanisms for detecting and mitigating malicious behaviors, ensuring the integrity and reliability of the inference process.

\section{Related Work}
Various approaches have been proposed for trustless AI inference in decentralized environments. Each method offers unique advantages but also has significant drawbacks, particularly when applied to large artificial neural networks (ANNs).

\paragraph{Proof of Quality (PoQ).}
PoQ ensures robust and efficient deployment of generative AI models on blockchain architecture by focusing on validating model responses using lightweight, simple model assessors. However, the trade-off between guaranteed accuracy and inference latency is significant, leading to less than 70 percent accuracy for fast inference, making it less suitable for quality-demanding applications \cite{poq2024}.

\paragraph{Zero-Knowledge Machine Learning (ZKML).}
ZKML combines zero-knowledge proofs (ZKPs) with machine learning for verifiable AI model inferences. Despite its strong guarantees, ZKML's computational overhead can be significant, limiting its practicality for large-scale applications \cite{zkml2023}.

\paragraph{Halo2 ZK-SNARK Protocol.}
The Halo2 protocol employs Plonkish arithmetization to efficiently express DNN inference and generate zero-knowledge proofs. While it ensures high security at lower costs than ZKML, the computational expenses are still prohibitive for very large neural networks \cite{halo22024}.

\paragraph{Optimistic Machine Learning (OPML).}
OPML introduces a novel framework that leverages optimistic rollups for machine learning inferences on the blockchain, significantly reducing computational costs and improving efficiency. Unlike ZKML, which uses zero-knowledge proofs to achieve verifiable AI inferences, OPML relies on a fraud-proof protocol that separates execution from proving, thereby optimizing resource usage while maintaining security. However, this trade-off comes with the need for a challenge period, which may delay finality compared to instant ZKP verification \cite{opml2024}.

\paragraph{Federated Learning.}
Federated learning enables decentralized training by aggregating model updates instead of raw data. However, communication overhead and synchronization challenges can slow down the overall process in heterogeneous environments \cite{fedlearn2024}.

\paragraph{Blockchain-based Model Verification.}
Smart contracts on blockchains can verify AI model inference by checking the consistency of model outputs. The execution of verification algorithms can be slow and resource-intensive, hindering real-time inference \cite{blockchain2024}.

\paragraph{Enclaves and Trusted Execution Environments (TEEs).}
TEEs, such as Intel SGX and ARM TrustZone, provide secure environments for AI model inference. Their limited scalability and availability overhead make them less suitable for large-scale, real-time applications \cite{tee2024}.

\paragraph{Homomorphic Encryption.}
Fully Homomorphic Encryption (FHE) allows computations on encrypted data, preserving privacy and validity. However, the significant performance overhead and resource requirements limit its practicality for real-time AI inference \cite{fhe2023}.

\paragraph{Verifiable Computation.}
Interactive proofs and Probabilistically Checkable Proofs (PCPs) enable verification of computations in a decentralized manner. The complexity and resource intensity of these methods limit their practicality for large-scale neural networks \cite{vc2023}.

\section{Self-Supervised Agents}
Collaborative agent-based frameworks have demonstrated strong results in improving both the accuracy and efficiency of AI systems, particularly when compared to monolithic solutions, by distributing tasks across specialized agents \cite{shi2024}. Our approach leverages the collective intelligence of swarms of agents capable of both inference and ranking. These agents form dynamic networks that ensure high-quality responses while effectively mitigating the risk of malicious behavior. In this section, we describe the architecture, practical parameters, and methods for detecting and mitigating malicious attacks.

\subsection{Agent Architecture}
Each agent in the swarm performs data inference and quality ranking. Agents communicate with each other to form a \textit{consensus} on the best responses, ensuring robustness and accuracy.

There is a valid question whether we can use the single LLM both for content generating and response ranking. Recent studies \cite{rrescue2023} answer positively to this question, moreover \textbf{MetaRanking} approach achieves over 80 percent accuracy with ranking complex GPT4 responses using lightweight Phi2 model\cite{metaranking2024}. Thus, each agent can easily provide reasonable ranking capabilities to the swarm using its lightweight expert LLM.

The proposed system employs a multi-agent architecture, where each agent is designed as a composite entity encompassing several key components:

\begin{enumerate}
    \item \textbf{Primary Cognitive Module:} This core component is responsible for both content generation and ranking tasks. It can be implemented as either:
    \begin{itemize}
        \item A Large Language Model (LLM), or
        \item An Expert System
    \end{itemize}
    
    \item \textbf{Auxiliary Processing Unit (Optional):} This module augments the primary cognitive capabilities by performing:
    \begin{itemize}
        \item Pre-processing operations on input data
        \item Post-processing operations on generated content
    \end{itemize}
    Auxillary Unit can connect to the external world and be represented like tools (interpreters, calculators, knowledge bases, filters) as well as third-party services (Math processors, Internet access and Search Engine, Cloud ML Providers and Co-Pilots)
\end{enumerate}

This modular architecture allows flexibility in agent implementation while maintaining a standardized interface for inter-agent communication and system integration.

The primary cognitive module's dual functionality in content generation and ranking enables efficient task execution, while the optional auxiliary processing unit provides enhanced adaptability to diverse input and output requirements.

\FloatBarrier
\begin{figure}[H]
    \centering
    \includegraphics[width=0.99\linewidth]{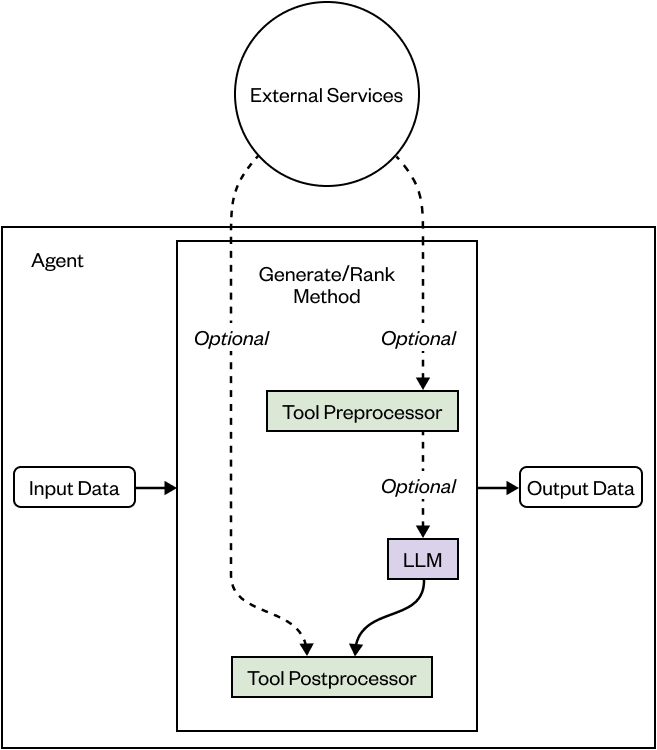}
    \caption{Schematic representation of the Agent architecture}
    \label{fig:agent_architecture}
\end{figure}
\FloatBarrier

\subsection{Swarm-based Consensus Mechanism for Optimal Response Selection}
This section outlines a novel swarm-based consensus mechanism designed to coordinate agents and select the most appropriate response from a pool of generated answers. The process is divided into three main phases: response generation, selective ranking, and final selection.
\paragraph{Response Generation Phase}
The process initiates with the following steps:

A client request is broadcast to the swarm.
Participating agents generate and submit encrypted responses to the swarm. It's important to note that only willing agents can take part in the response generation, filtering client requests for their expertise and applicability. After a predetermined time interval, $\Delta t$, the responses are decrypted using keys submitted by each participating agent.

\begin{equation}
R_i = E(r_i, k_i), \quad D(R_i, k_i) = r_i
\end{equation}
where $R_i$ is the encrypted response, $r_i$ is the original response, $k_i$ is the encryption key, $E$ is the encryption function, and $D$ is the decryption function.
This encryption-decryption mechanism mitigates potential attacks involving the copying of highly-rated nodes' responses.
\paragraph{Selective Ranking Phase}
Following the response generation, a selective ranking process is employed:

A subset of agents, $S_j \subset A$, is pseudo-randomly selected to rank each agent's response, where $A$ is the set of participating agents.
The selection is based on a recent block hash, $H_b$, ensuring randomness and fairness.
Each agent $a_i \in A$ ranks responses from approximately one-third of other agents, excluding its own response.

\begin{equation}
S_j = f(H_b, A \setminus {a_j}), \quad |S_j| \approx \frac{|A| - 1}{3}
\end{equation}
where $f$ is the selection function based on the block hash.
This selective ranking approach reduces the risk of collusion and prevents any single agent from exerting undue influence over the consensus.
\paragraph{Final Selection Phase}
The final selection of the optimal response proceeds as follows:

Rankings are submitted to the swarm in encrypted form to prevent copying attacks.
After a decryption round, the best response is selected based on weighted rankings.
The weight of each ranking is determined by the rating of the ranking agent.

\begin{equation}
r_{best} = \underset{r_i}{\operatorname{arg,max}} \sum_{j \in S_i} w_j \cdot \operatorname{rank}(r_i, a_j)
\end{equation}
where $w_j$ is the weight (rating) of agent $a_j$, and $rank(r_i, a_j)$ is the rank assigned to response $r_i$ by agent $a_j$.

The selected best response, $r_{best}$, is returned to the client.

This mechanism ensures a fair, decentralized, and robust selection process that leverages the collective intelligence of the swarm while mitigating potential vulnerabilities.

\FloatBarrier
\begin{figure}[H]
    \centering
    \includegraphics[width=0.99\linewidth]{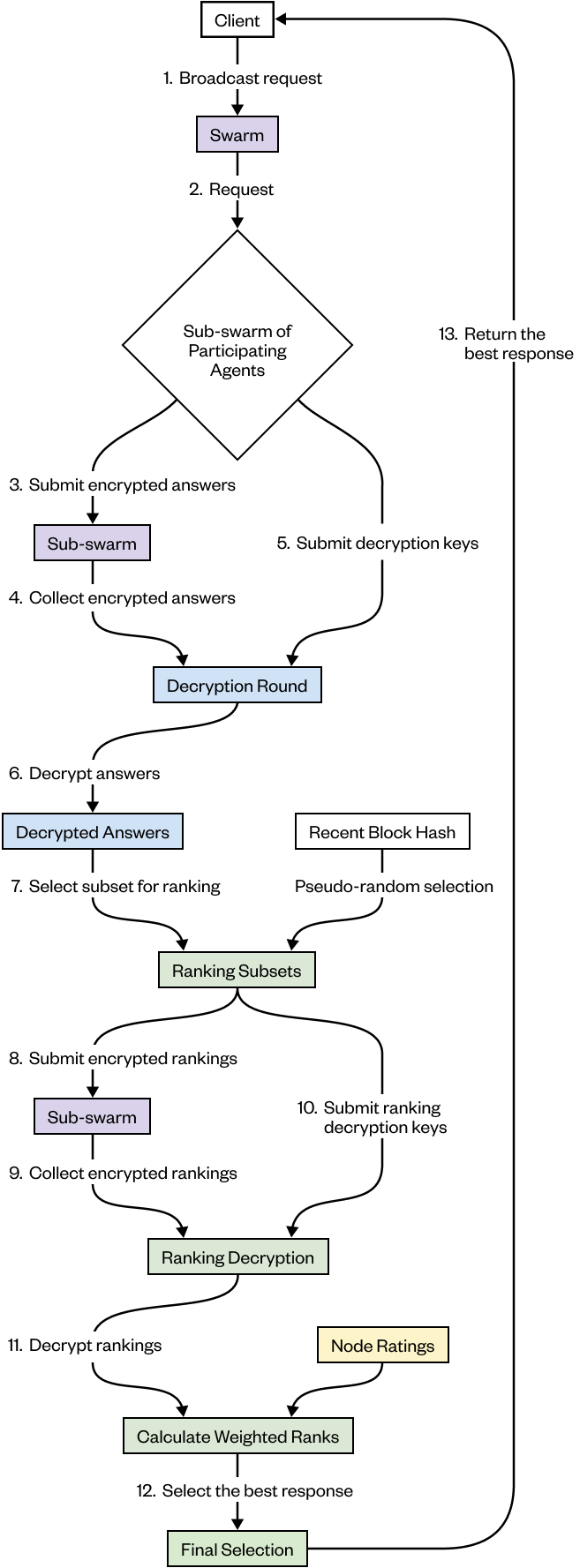}
    \caption{Swarm-based Consensus Mechanism}
    \label{fig:response_flow}
\end{figure}
\FloatBarrier

\subsection{Agents Rating - Ranking Ability and Quality Estimation}
Trustless systems need to be robust and stable with the highest amount of noise being introduced by malicious or lazy actors. While the smarm's ranking algorithm guarantees consensus with enough swarm size, an effective agent ranking ability and answer quality estimation enhance these mechanisms by ensuring that only the most reliable nodes participate in the consensus process. Our approach leverages the statistical properties of score deviations from the mean, assuming that the collective rankings by nodes conform to a normal distribution, a consequence of the Central Limit Theorem.

\subsubsection{Methodology}
Each agent in the network ranks other agents' contributions based on specific criteria. For each ranking cycle, the deviation of a node's score from the mean score is computed. The standard deviation of these score deviations across multiple cycles is used as a metric to estimate the node's ranking ability:

\begin{equation}
    \sigma_i = \sqrt{\frac{1}{N-1} \sum_{j=1}^{N} (x_{ij} - \overline{x}_j)^2}
\end{equation}

where $\sigma_i$ is the standard deviation of score deviations for node $i$, $x_{ij}$ is the score given by node $i$ to node $j$, and $\overline{x}_j$ is the mean score received by node $j$.

Limiting this metric only to the blocks with a high number of participants, we assume that the sum of scores given to each node approximates a normal distribution (the consequence of the Central Limit Theorem). This assumption allows us to use statistical methods to systematically analyze the ranking behaviors.

Nodes with lower values of $\sigma_i$ are considered more aligned with the collective decision-making process, indicating a higher reliability in their ranking assessments or response quality. These nodes are given more influence in the consensus process, leading to a more robust and secure agent swarm.

Finally, to estimate the \textit{node's rating} (ranking ability) we can calculate the inverted $\sigma_i$ value:

\begin{equation}
     Rating_i = 1 - \sigma_i
\end{equation}

\subsubsection{Nodes' Rating Estimation}
We conducted several simulations to estimate ranking ability. In our setup, we used 10 test agents with different actual ranking abilities, and we simulated various numbers of test consensus rounds to calculate the proposed ranking metric. As we observed, even 10 simulation rounds are sufficient to provide a rough estimation of an agent’s ranking ability.

\FloatBarrier
\begin{figure}[H]
    \centering
    \includegraphics[width=0.99\linewidth]{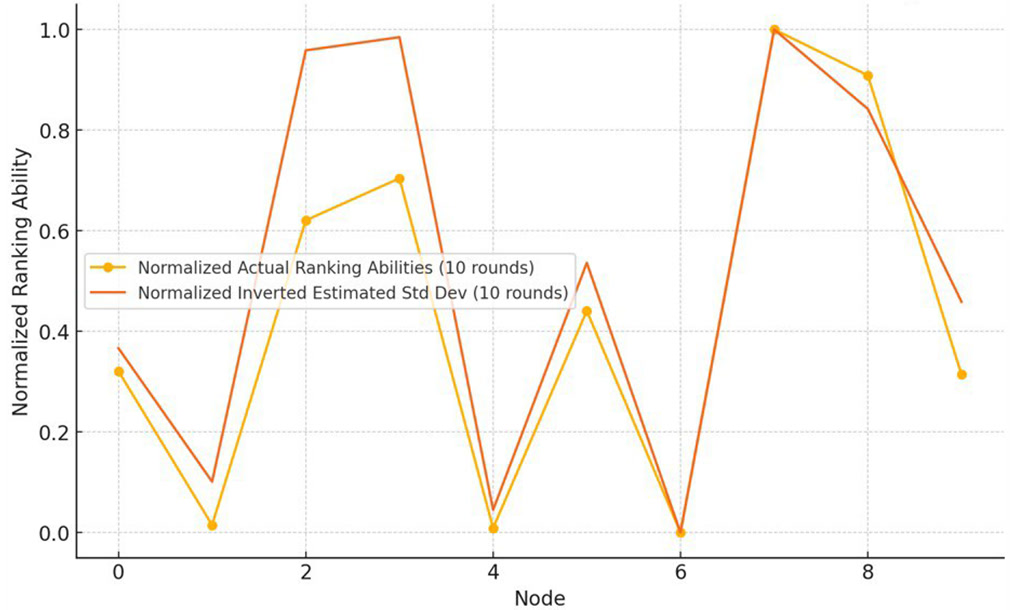}
    \caption{Ranking Estimation Simulation - 10 Rounds}
    \label{fig:ranker_est_10}
\end{figure}
\FloatBarrier

\FloatBarrier
\begin{figure}[H]
    \centering
    \includegraphics[width=0.99\linewidth]{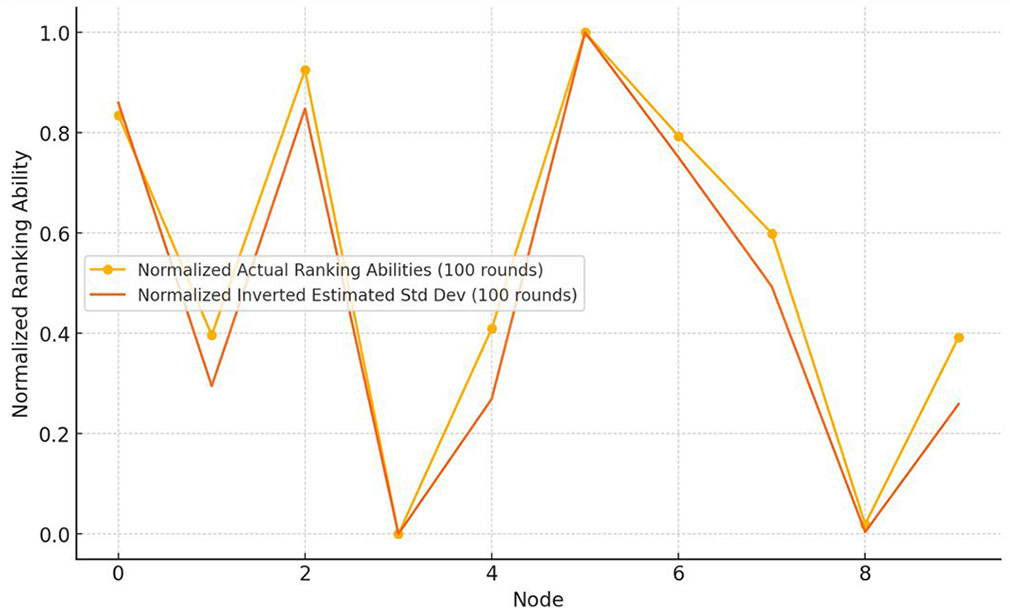}
    \caption{Ranking Estimation Simulation - 100 Rounds}
    \label{fig:ranker_est_100}
\end{figure}
\FloatBarrier

\FloatBarrier
\begin{figure}[H]
    \centering
    \includegraphics[width=0.99\linewidth]{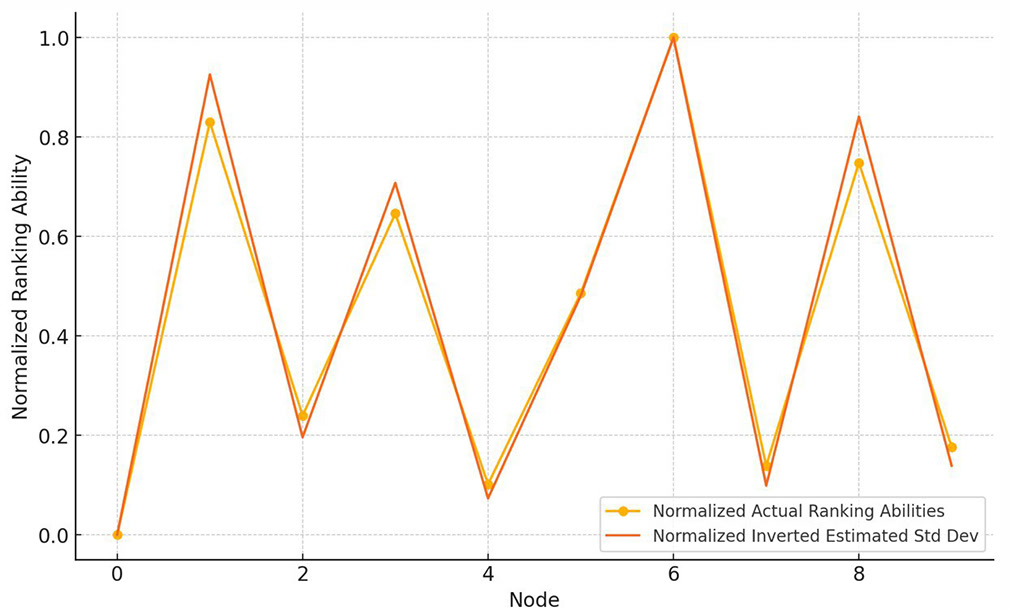}
    \caption{Ranking Estimation Simulation - 1000 Rounds}
    \label{fig:ranker_est_1000}
\end{figure}
\FloatBarrier

This ranking estimation mechanism provides a statistically sound method for assessing the reliability and accuracy of agents in consensus. By integrating this system, agent swarm enhances their security and efficiency, ensuring that only the most competent nodes govern the consensus process.

\section{Adversarial Agent Detection and Mitigation}
We model various types of malicious attacks, including those producing random outputs (lazy agents), inconsistent quality outputs (buggy agents), and deliberate attempts by malicious agents to forge outputs for personal gain. While lazy and buggy agents' outputs are mitigated using ranking and rating systems, our approach includes mechanisms for detecting and isolating malicious agents' influence on the cluster, ensuring the integrity and reliability of the inference process.

\subsection{Sybil Attacks}
A Sybil attack poses a significant security threat in trustless networks, where a single entity creates multiple fake identities to gain disproportionate influence or control. A Self-Supervised Agent Inference approach necessitates a balanced incentive model to render Sybil attacks financially unfeasible.

\subsubsection{Methodology}
Each participating node in the network is tasked with completing a specific job, which involves computing LLM requests. This job completion is essential as it directly influences the consensus process by providing a measurable output that can be verified and ranked by other nodes.

Nodes that wish to participate in the solution process are required to purchase a ticket. This economic disincentive is a critical component in preventing Sybil attacks, as the cost of creating numerous fake identities becomes prohibitively high due to the required token price for each participating identity.

The reward mechanism in this consensus model is designed to promote the best contributions. Only the node whose solution wins the majority approval from the ranked nodes receives the major portion of the rewards, which include both the intrinsic value derived from solving the problem (e.g., transaction fees, block rewards) and part of the ticket value from nodes that offered poorer solutions. This further aligns the incentives of the nodes with the overall health and security of the network.

\subsubsection{Sybil Attack Profitability Simulation}
The following graph displays the profitability of Sybil attacks in an agent swarm, with varying numbers of total nodes and token deposit requirements. The x-axis represents the total number of nodes in the network, ranging from 10 to 500. The y-axis shows the range of token deposits required by each node, from 0 to 2 tokens. In this scenario, we simulate a reward of 20 tokens for the winning node.

\FloatBarrier
\begin{figure}[H]
    \centering
    \includegraphics[width=0.99\linewidth]{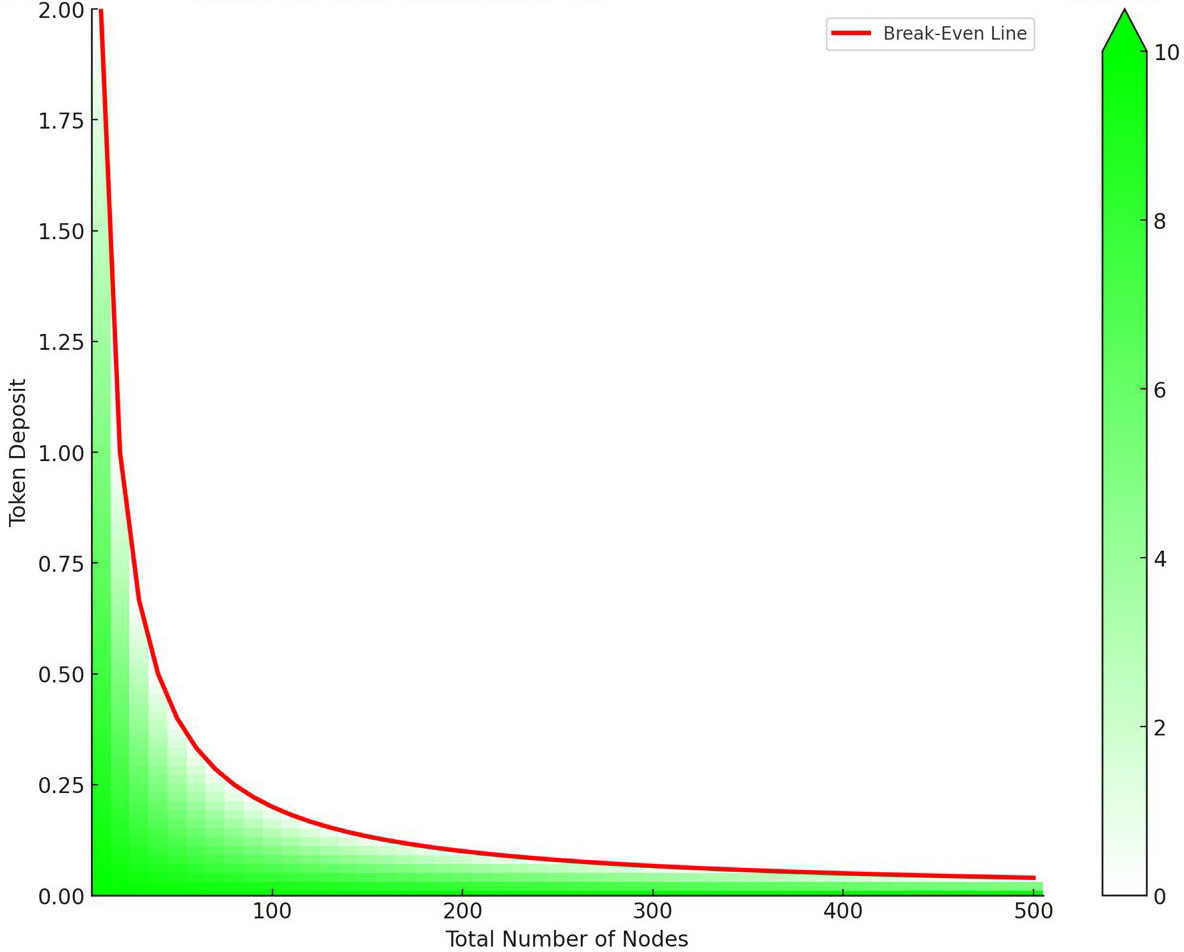}
    \caption{Profitability of Sybil attacks}
    \label{fig:sybil_attack_reward_chart}
\end{figure}
\FloatBarrier

The color coding indicates the financial outcome of the Sybil attack:

\begin{itemize}
    \item \textbf{White areas} represent scenarios where the attack results in negative profits, making them unprofitable or unattractive for attackers.
    \item \textbf{Shades of green} signify varying levels of positive profits, with darker greens indicating higher profits from the attack.
\end{itemize}

A red line on the graph marks the break-even point, where the profits from the attack shift from positive to zero. Above this line, attacks are not profitable (white area), providing a clear visual guide on setting token deposit thresholds to deter malicious activities effectively. This line is crucial to understanding how to scale security measures based on network size and potential rewards to maintain the integrity of the network.

In summary, the graph demonstrates that setting a ticket price as low as 1\% of the potential reward is sufficient to make an attack unprofitable, even in a moderately sized agent swarm, effectively enhancing the network's resilience against Sybil attacks.

The selective ranking based on hash functions reduces the overhead typically associated with each node evaluating every other node’s submission, which can be particularly burdensome in large networks. Furthermore, by decentralizing the ranking and not allowing nodes to rank their own solutions, the system inherently guards against self-promotion and favoritism, thus enhancing the security and integrity of the consensus process.

This consensus mechanism leverages economic incentives and algorithmic randomness to create a robust defense against Sybil attacks, ensuring that the blockchain maintains its decentralized, secure, and transparent nature.

\subsection{Prompt Engineering Attacks}
 One of the potential vulnerabilities in the swarm of agents approach is the prompt engineering attacks. These attacks occur when malicious nodes attempt to manipulate the inference process by crafting inputs (prompts) that can exploit the ranking mechanism of other agents in the network. The goal of such an attack is to have the malicious agent’s output ranked higher than it deserves, thereby influencing the overall consensus of the swarm and potentially compromising the system’s integrity.

Prompt engineering attacks can take various forms, but they generally fall into two main categories: low-frequency token attacks and common-sense prompt attacks. Both types of attacks aim to deceive or manipulate the underlying language models (LLMs) used by other agents to skew the ranking of responses in favor of the attacker.

\subsubsection{Low-Frequency Token Attacks}
Low-frequency token attacks involve the insertion of rare or special tokens, such as unique Unicode characters, into the generated text by the malicious agent. The premise behind this attack is that by introducing these low-frequency tokens, the malicious response might exploit biases or vulnerabilities in the language models of other agents, causing them to rank the manipulated response more favorably. This tactic could potentially work in scenarios where the language models disproportionately weight the presence of rare tokens as indicative of novel or important information.

However, a key countermeasure against such attacks in our decentralized AI framework is the heterogeneity of the agents' LLMs. Since each agent in the swarm operates either a varied or uniquely tuned LLM, the effectiveness of low-frequency token attacks is significantly diminished. An insertion that may trigger a favorable bias in one model is unlikely to have the same effect across a diverse set of models. The diversity of LLM architectures, training data, and tokenization processes means that these rare tokens do not consistently influence the ranking process, thereby reducing the chances of a successful attack. This diversity acts as a natural defense mechanism, ensuring that no single token-based strategy can universally deceive the swarm.

\subsubsection{Common-Sense Prompt Attacks}
Common-sense prompt attacks are another form of manipulation where the malicious agent embeds statements within the prompt that are intended to influence the other agents’ LLMs to rank their response higher. Examples of such manipulative statements include phrases like “this answer is the best” or other similar assertions that aim to exploit basic common-sense reasoning or self-referential biases within the LLMs.

Each agent in the swarm is inherently motivated to protect itself against manipulations to maximize its received incentive. Luckily such a self-referential protection mechanism is grounded in the “common-sense” capabilities of the agents’ LLMs, and can be utilized to penalize responses that attempt to unduly influence ranking through non-substantive means.

\section{Evaluation}
We estimate the computational resources required for our approach, considering factors such as network latency, processing power, and data throughput. Our model shows that a swarm of agents can operate very efficiently in a decentralized environment.

\paragraph{Inference Latency Comparison.}
Table \ref{tab:latency-comparison} compares the inference latencies of various decentralized AI approaches, revealing significant differences in performance. Proof of Quality (PoQ) demonstrates the lowest latency at 50 ms for MobileNet v2, though it is achieved with low accuracy guarantees ($<$ 70$\%$) due to validation using a simple BERT transformer   \cite{poq2024}. In stark contrast, ZKML exhibits extremely high latency, taking over 24 hours for ResNet-50 inference, rendering it impractical for large-scale models \cite{zkml2023}. The Halo2 ZK-SNARK protocol shows improvement but still requires 2457.5 seconds (about 41 minutes) for MobileNet v2, which remains prohibitive for very large neural networks \cite{halo22024}.
OPML offers a significant reduction compared to ZKML, completing ResNet-50 inference in 3.6 hours, due to its challenge period \cite{opml2024}. Trusted Execution Environments like Intel SGX achieve relatively low latency at 230 ms for VGG-16, but struggle with scalability \cite{tee2024}. Homomorphic Encryption, while preserving privacy, incurs high latency at 788 seconds for SqueezeNet, making it unsuitable for real-time applications \cite{fhe2023}.
Federated Learning, Blockchain-based Verification, and Verifiable Computation approaches lack specific latency figures in the literature for direct comparison, but are generally understood to face significant overhead due to communication costs, resource-intensive operations, and computational complexity, respectively \cite{fedlearn2024,blockchain2024,vc2023}.
This comparison highlights the ongoing challenge in achieving both low-latency and secure decentralized AI inference. While some approaches like PoQ and TEEs offer promising latency figures, they come with their own limitations. The trade-off between security, efficiency, and practicality remains a key area for further research and development in decentralized AI, and Self-Supervised Agent Inference shows a promising approach to this problem.

\paragraph{Ultra-Low Latency Inference on Large Language Models.}
Our swarm-based consensus mechanism demonstrates remarkable efficiency when applied to large language models such as Llama 3 405B, achieving inference latencies of under 125 milliseconds. This exceptional performance is attributed to several key factors:

\begin{itemize}
    \item \textit{Parallel Processing:} The swarm architecture allows for massively parallel response generation. All agents taking part in the round can simultaneously process the input query, effectively avoiding latency on the initial inference step.
    \item \textit{Selective Ranking:} By having each agent rank only a subset of responses, we drastically reduce the time required for the evaluation phase without compromising the quality of selection.
    \item \textit{Rapid Ranking Process:} To rank a response, only a single token needs to be inferred, which is orders of magnitude faster than generating a full response. This allows for quick and efficient quality assessment (ranking) without significantly impacting overall latency.
    \item \textit{Weighted Ranking Aggregation:} The final selection phase employs a computationally efficient weighted ranking system that quickly identifies the optimal response.
    \item \textit{Asynchronous Operations:} Many of the consensus mechanism's steps occur asynchronously, allowing for overlapping operations that further reduce overall latency.
\end{itemize}

This combination of parallelism, efficient cryptographic operations, intelligent agent coordination, and our consensus technique enables our system to leverage the full power of the Llama 3 405B model while maintaining inference latencies below 125 ms. The single-token ranking approach is particularly crucial, as it allows for quality assessment at a fraction of the time required for full response generation. This represents a significant advancement in decentralized AI, offering performance comparable to centralized solutions while preserving the benefits of swarm-based consensus and decentralization.

\begin{table*}[t]
\centering
\caption{Comparison of Inference Latency for Decentralized AI Approaches}
\label{tab:latency-comparison}
\begin{tabular}{lp{2cm}ll}
\hline
\textbf{Approach} & \textbf{Latency} & \textbf{Model/Dataset} & \textbf{Limitations} \\
\hline \\
PoQ & 50 ms & MobileNet v2 & Low inference accuracy guarantees ($<$ 70$\%$) \\
ZKML & $>$24 h & ResNet-50 & Impractical at scale \\
ZK-SNARK Halo2 & 2457.5 s & MobileNet v2 & Prohibitive for large nets \\
OPML & 3.6 h & ResNet-50 & Challenge period delays \\
Federated Learning & N/A & -- & Communication overhead \\
Blockchain Verification & N/A & -- & Resource-intensive \\
TEEs (SGX) & 230 ms & VGG-16 & Limited scalability \\
HE & 788 s & SqueezeNet & Not real-time capable \\
Verifiable Computation & N/A & -- & Small networks only \\
\hline \\
\textbf{Ours} & \textbf{$<$ 125 ms} & Llama 3 405B & -- \\
\hline \\
\end{tabular}
\end{table*}

\section{Conclusion}
This paper presents a novel AI inference approach that leverages the collective intelligence of agent swarms in decentralized environments. By employing agents capable of data inference and response ranking, our method enables Large Language Models (LLMs) to function effectively as response classifiers while addressing critical challenges including security, reliability, and rapid response times.

Our proposed method demonstrates significant improvements over existing trustless inference strategies in terms of speed, accuracy, and resilience to malicious attacks. By modeling various types of malicious agent behavior, we have developed and validated mechanisms to detect and mitigate such threats, ensuring the integrity of the inference process. This approach provides a scalable solution adaptable to the growing demands of AI applications.

The methodologies and insights presented here suggest that decentralized AI systems have the potential to match or surpass the performance of traditional centralized systems. Our work represents a significant step forward in trustless and scalable AI inference, highlighting the potential of swarm-based intelligence to overcome inherent challenges in these environments.

Future research may build upon this work to further improve response quality and accuracy through more complex inter-agent collaboration, and advancement of heterogeneous agent networks. Additionally, efforts could be directed towards optimizing scalability and exploring applications in diverse, complex real-world scenarios. These advancements will pave the way for more secure, efficient, and reliable AI applications across various domains, contributing to a new paradigm of decentralized and permissionless inference.

\begin{small}

\end{small}

\begin{thebibliography}{99}

\bibitem{sakshi2024}
SAKSHI. "Decentralized AI Platforms." arXiv preprint arXiv:2307.16562, 2024.

\bibitem{pocket2024}
POKT Network. "Decentralized AI: Permissionless LLM Inference on POKT Network." arXiv preprint arXiv:2405.20450, 2024.

\bibitem{aiprotocol2024}
AI Protocol. "Decentralized Inference Clusters." AI Protocol Whitepaper, 2024.

\bibitem{nesa2024}
 Nesa AI. "Introduction to Nesa." Nesa Documentation, 2024.
 
\bibitem{poq2024}
Zhang, Zhenjie, Rao, Yuyang, Xiao, Hao, Xiao, Xiaokui, Yang, Yin. "Proof of Quality: A Costless Paradigm for Trustless Generative AI Model Inference on Blockchains." arXiv preprint arXiv:2405.17934, 2024.

\bibitem{zkml2023}
Feng, B., Qin, L., Zhang, Z., Ding, Y., and Chu, S. Zen. "An optimizing compiler for verifiable, zero-knowledge neural network inferences." Cryptology ePrint Archive, 2021.

\bibitem{halo22024}
Daniel Kang, Tatsunori Hashimoto, Ion Stoica, Yi Sun. "Scaling up Trustless DNN Inference with Zero-Knowledge Proofs." arXiv preprint arXiv:2210.08674, 2024.

\bibitem{opml2024}
KD Conway, Cathie So, Xiaohang Yu, Kartin Wong. "opML: Optimistic Machine Learning on Blockchain." arXiv preprint arXiv:2401.17555, 2024.

\bibitem{fedlearn2024}
J. Wang, A. K. Sahu, G. Joshi, and S. Kar, “Matcha: A matchingbased link scheduling strategy to speed up distributed optimization,”
IEEE Trans. Signal Process., vol. 70, pp. 5208–5221, Nov. 2022.

\bibitem{blockchain2024}
Sanghyeon Park, Junmo Lee, Soo-Mook Moon. "Blockchain-based Model Verification for AI Inference." arXiv preprint arXiv:2305.04062, 2024.

\bibitem{shi2024}
Yunxiao Shi, Min Xu, Haimin Zhang, Xing Zi, and Qiang Wu. "A Learnable Agent Collaboration Network Framework for Personalized Multimodal AI Search Engine." arXiv preprint arXiv:2409.00636, 2024.

\bibitem{tee2024}
Jean-Baptiste Truong, William Gallagher, Tian Guo, Robert J. Walls. "Memory-Efficient Deep Learning Inference in Trusted Execution Environments" arXiv preprint arXiv:2104.15109, 2021.

\bibitem{fhe2023}
Qian Lou, Lei Jiang. "HEMET: A Homomorphic-Encryption-Friendly Privacy-Preserving Mobile Neural Network Architecture" arXiv preprint arXiv:2106.00038, 2021.

\bibitem{vc2023}
Tiantian Gong, Aniket Kate, Alexandros Psomas, Athina Terzoglou. "V3rified: Revelation vs Non-Revelation Mechanisms for Decentralized Verifiable Computation" arXiv preprint arXiv:2408.07177, 2024.

\bibitem{rrescue2023}
Zhang, Y., Sun, Q., Xu, Z., et al. "RRescue: Ranking LLM Responses to Enhance Reasoning Over Context." arXiv preprint arXiv:2311.09136, 2023.

\bibitem{metaranking2024}
Wang, H., Zhu, Y., Lin, Z., et al. "Meta Ranking: Less Capable Language Models are Capable for Single Response Judgement." arXiv preprint arXiv:2402.12146, 2024.

\end{thebibliography}
\end{document}